\documentclass[12pt]{elsarticle}

\usepackage{amssymb}
\usepackage{lineno}
\usepackage{cite}
\usepackage{graphicx}  
\usepackage{amsmath}
\usepackage[ruled]{algorithm2e}

\usepackage{tikz}	   
\usepackage{tikzscale}
\usepackage{pgfplots}
\pgfplotsset{compat=1.12}

\usepackage{listings}
\journal{Journal of Systems and Software}
\begin{document}
\begin{frontmatter}
\title{Systematic Testing of Genetic Algorithms: A Metamorphic Testing based Approach}

\author{Janette Rounds and Upulee Kanewala}
\address{Gianforte School of Computing\\
Montana State University\\
Bozeman, Montana 59717\\
Email: janette.rounds@msu.montana.edu\\
Email: upulee.kanewala@montana.edu}

\begin{abstract} 
Genetic Algorithms are a popular set of optimization algorithms often used to aid software testing. However, no work has been done to apply systematic software testing techniques to genetic algorithms because of the stochasticity and the lack of known outputs for genetic algorithms. Statistical metamorphic testing is a useful technique for testing programs when the output is unknown or when the program has random elements. In this paper, we identify 17 metamorphic relations for testing a genetic algorithm and show, through mutation testing, that these relations are more effective at finding defects than traditional unit tests based on known outputs.  We examined the failure rates of the system-level relations when initialized with various fitness functions. We found three relations failed excessively and we then modified these relations so that they failed less often. We also identified some metamorphic relations for genetic algorithms that are generalizable across different types of evolutionary algorithms and showed that our relations had similar mutation scores between two implementations.  This is the first time statistical metamorphic testing has been applied for testing genetic algorithms. 

\end{abstract}

\end{frontmatter}
\section{Introduction}
\label{sec:intro}
Genetic Algorithms are a popular optimization technique that have been used in domains from consumer goods quality assurance \citep{lee2016slippery} and modeling traffic flows \citep{chia2016traffic} to medical imaging \citep{pareek2016medical} and materials science \citep{davis2016application}. Additionally, Genetic Algorithms are often used to optimize the parameters or structure of other machine learning algorithms \citep{perreault2015swarm}. There has been a significant amount of work done on using genetic algorithms to test software \citep{structural1996genetic,wegener1997testing,rao2015genetic} especially on using genetic algorithms in mutation testing. However, there is nothing in the literature about testing the genetic algorithms themselves even though they are often used for critical applications such as those mentioned above. This is especially problematic given that genetic algorithm's are being used as a tool to help medical diagnoses \citep{tan2003evolutionary,pena1999fuzzy} and in other high-stakes situations. 

There has been little work on testing machine learning algorithms in general \citep{xie2009application}. The reasons why so little work has been done in this area are myriad. The first reason is that most machine learning algorithms incorporate at least one random element. Random elements are challenging to test because the same input can produce different results \citep{guderlei2007statistical}. 

Furthermore, the correct answer is often unknown. For example, a person can design a genetic algorithm to optimize land zoning subject to multiple constraints \citep{stewart2004genetic}. In that instance, if the true answer was known, the land would already have been zoned. If the true answer is in fact known, like many of the optimization problems we discuss here, the true answer and the answer a genetic algorithm returns may not match, even if the genetic algorithm is correct.  For example, in the Ackley's function which we discuss later in this paper, a genetic algorithm may not find the true global minimum for the function at (0,...,0) for a number of dimensions. Instead the genetic algorithm may find one of the many local minima for the Ackleys function and return this value. Although this local minimum may not be the true answer, the genetic algorithm still may be correct. 

Genetic algorithms are useful tools for approximation. However, they tend not to return exact answers, but an approximation of a good answer. If a genetic algorithm returned an exact and correct answer for a problem in which the answer was known, especially if it returned the same answer several times, we would probably suspect the genetic algorithm of having a coding error. 

To test genetic algorithms we can use a testing technique called metamorphic testing, first proposed by Chen et. al. \citep{chen1998metamorphic}. Metamorphic testing involves defining properties by which we can relate two or more outputs from an algorithm based on the input. If the outputs of two related inputs do not follow the property, there is an error in the program under test.  Statistical metamorphic testing \citep{guderlei2007statistical} is a useful technique when the program under test has random elements. Instead of an initial and follow-up test case, statistical metamorphic testing uses an initial and follow-up sample, which is then tested using statistical hypothesis testing. 

We identified 17 metamorphic relations from the literature on genetic algorithms. We then used mutation testing to demonstrate the effectiveness of the various relations on two different implementations. We also demonstrated the effectiveness of three of these relations on two different differential evolution implementations. We found that traditional deterministic unit tests were not as effective at finding mutations as metamorphic relations, and that system-level relations and unit-level relations perform well on different parts of the genetic algorithm. We also found that our three relations for differential evolution performed as well or better than the twelve relations for genetic algorithms. 

The rest of this paper is organized as follows: Section \ref{sec:relwork} contains related work and background on metamorphic testing, genetic algorithms, and mutation testing. In Section \ref{sec:imple}, we discuss our implementation of a genetic algorithm. In Section \ref{sec:metarel}, we identify 17 metamorphic relations, five for fitness functions, three for genetic algorithm operators, and nine system-level level relations. In Section \ref{sec:exp}, we lay out several experiments we conducted. Finally, Section \ref{sec:concl} contains our conclusions and future work. 

\section{Background and Related Work}
\label{sec:relwork}
Genetic algorithms are often used in software testing. The most common uses of genetic algorithms are in test case generation \citep{geronimo2012parallel}, multi-objective test case generation \citep{henard2013multi}, test case prioritization \citep{sharma2014applying}, and mutation testing \citep{rao2015genetic}. There has been no work published in the literature on how to test genetic algorithms. However, Arcuri and Briand \citep{arcuri2014hitchhiker} state that many randomized algorithms (like a genetic algorithm) are used in software applications. Additionally, Arcuri and Briand argue for the use of statistical testing in software applications that use random algorithms. 

Metamorphic testing is a testing technique used when the algorithm is non-deterministic and/or when there is no way to determine the correct output \citep{chen1998metamorphic}. To conduct metamorphic testing, first one defines a metamorphic relation (or a set of metamorphic relations) for the program under test. A metamorphic relation is a property by which we can relate two outputs based on the input.  Then, tests are defined for each metamorphic relation. Each test consists of an initial test case and a follow-up test case based on the metamorphic relation. Tests are then executed on the program under test. The output of the initial and follow-up test cases are evaluated to determine if the text cases follow the metamorphic relation.

When the algorithm is stochastic, one cannot simply check if the result of the follow-up test is equal to the expected value. The result may be close, but because of the stochasticity, the result will not be exactly equal. Therefore, statistical tests are applied to determine whether the difference between the expected output and the actual output is statistically significant. This is called statistical a relationship between two samples, an initial and a follow-up test sample, is specified, usually taking the form of null and alternative hypotheses. The output of the program generates the samples and is compared using statistical hypothesis testing. 

Genetic Algorithms all consist of an encoding of potential solutions as chromosomes, e.g. a bit string, and a fitness function \citep{mitchell1997machine}, which is a function that evaluates each potential solution and returns a value based on that evaluation. A generic genetic algorithm has a population of properly encoded potential solutions which is usually randomly generated, and a randomized way or ways to change potential solutions to see if the fitness of that solution improves. Unfit solutions are removed from the population. 

Testing a genetic algorithm presents several problems. Genetic algorithms are randomized algorithms. Each time the genetic algorithm is run, the output of the algorithm will be different. There is also no way to determine the "correct" output of a genetic algorithm. In many cases, if the target output of a genetic algorithm is known, there is often no reason to use a genetic algorithm at all. Furthermore, if the data provided to the genetic algorithm is misleading in some way, the output of the genetic algorithm will not match the desired output. Thus, testing approaches that require one to compare an output against a true answer will not work for testing a genetic algorithm. 

In this paper, we use several problems where the correct answer is already known in order to demonstrate the process of testing a genetic algorithm. However, we also provide ways to expand the testing process in the case where the target answer is unknown. 

There has been no published research on how to test genetic algorithms. However, Xie et al. \citep{xie2011testing} defined several possible metamorphic relations to use when testing a machine learning classifier. In machine learning, classification is the problem of delineating decision boundaries so that all examples inside of a boundary are of one class. Genetic algorithms have been shown to be effective classifiers in a multitude of cases \citep{ishibuchi1997single,corcoran1994using}. Xie et al. \citep{xie2011testing} demonstrated how to test two other machine learning methods, the k-Nearest Neighbor and the Naive Bayes Classifiers, using metamorphic testing. Xie et al. also defined several possible metamorphic relations to use when testing a machine learning classifier. These authors showed that metamorphic testing is an effective way to test machine learning classifiers. However, they only tested this approach on WEKA \citep{hall2009weka}, an open source tool for performing classification, regression and other data mining tasks. 

Shin Yoo \citep{yoo2010metamorphic} showed how one could use metamorphic testing to validate a machine learning approach, called simulated annealing, to an optimization problem. Optimization problems consist of an objective function, $f(x)$ that we must either minimize or maximize subject to constraints. Shin Yoo showed that metamorphic relations can be an effective way to test machine learning approaches to optimization problems, especially for certain kinds of faults. Genetic algorithms are another way to approach an optimization problem. 

Murphy et.al. \citep{murphy2009automatic} showed how an automated metamorphic testing framework can be used to test support vector machines, decision trees, and ranking algorithms. The most complex algorithm tested by Murphy et. al. is the MartiRank algorithm. This algorithm is a type of ensemble method that divides the data into a series of sub-lists which it then orders according to the "quality" of the features, similar to a fitness function, except that it is iteration dependent. At each iteration, the model describes how to divide the data into lists and updates the quality measure. When all the rankings are completed, the algorithm reconstructs a final ranking based on the divisions and quality measures of previous iterations. Murphy et. al. designed an automated metamorphic testing system to improve the speed at which metamorphic tests can be developed. 

None of these papers test very complicated algorithms. The most complex of these is the MartiRank algorithm, which is based on a series of simple ranking algorithms. Genetic algorithms involve many more random components than even the most complex algorithm tested to date. In addition, Murphy et. al. and Xie et. al. used the Weka \citep{hall2009weka} implementation of these algorithms. There is no Weka implementation for a genetic algorithm. All of this means that there is a great need for testing genetic algorithms thoroughly and until now there has been no established way to accomplish this task.  

In order to demonstrate the effectiveness of the testing approach outlined in this paper, we will use mutation testing. Mutation testing is technique that has been shown to be effective for comparing testing techniques \citep{andrews2005mutation}. The first step in mutation testing is to generate a number of mutants, given the source code to the program under test. These mutants are identical to the program under test except that one line has been changed. Next, the test set is run on each mutant. If the test set detects the changed line (in other words, if at least one test fails with the changed line where the test passed without the line being changed), the mutant is said to have been 'killed'. If the test set does not detect the changed line, the mutant is said to have 'survived'. Compilation and run-time errors and tests that time out are considered 'killed' as well. Then, the number of killed mutants is counted and is divided by the total number of mutants. This gives a mutation score. A perfect mutation score would be 1, or all of the mutants detected by the test set. However, a perfect mutation score is usually impossible. Consider the example of a for loop in a Java program.
\begin{lstlisting}
for(int i = 0; i < 10; i++){}
\end{lstlisting}
If the code above is the original program and the following is the mutant:
\begin{lstlisting}
for(int i  = 0; i != 10; i++){}
\end{lstlisting} there would be no way to detect the mutant as it would be functionally equivalent to the original program. However, most mutants would not be functionally equivalent and could be detected by the test set, if an appropriate test exists. Therefore, the goal of mutation testing is to get the mutation score as close to 1 as is possible by changing tests so that they detect more mutants. In this paper, we compare the mutation score of different types of tests in order to determine which tests detect the most number of faults. 

\section{Genetic Algorithm Implementation}
\label{sec:imple}

Most often genetic algorithms used in classification are included as part of a larger solution. For example, a common use of genetic algorithms is to train the weights of a neural network. Genetic algorithms in an optimization context usually do not use any other algorithms. Additionally, it is trivial to turn a minimization problem into a maximization problem. This is important for the future generalization of the testing technique we develop here. Therefore we implemented a genetic algorithm that optimizes a function. 

\begin{algorithm}
    \SetKwInOut{Input}{Input}
    \SetKwInOut{Output}{Output}
    \Input{ $popSize$, $killRate$, $mutRate$, $\delta$, $maxGen$}
    \Output{the best solution from $P_{t}$}
	Randomly initialize population $(P_{0})$\;
	Evaluate fitness for $P_{0}$\;
	\While{$t < maxGen$ and best fitness $> \delta$}{
        $Child_{t}$ = selection $P_{t-1}$\;
        $Child_{t}'$ = crossover $Child_{t}$\;
        $Child_{t}''$ = mutate $Child_{t}'$\;
        Evaluate fitness for $Child_{t}''$\;
        $P_{t} = $ replace $P_{t-1}, Child_{t}''$\;
        $t += 1$\;
    }
    \caption{Genetic Algorithm}
    \label{algo:GA}
\end{algorithm}

A typical example of a genetic algorithm can be seen in Algorithm \ref{algo:GA}. The generic genetic algorithm takes as input a fitness function, a population size and a fraction of the population to be replaced at each time step, a mutation rate and a fitness threshold \citep{mitchell1997machine}. The output will be a set of real numbers that produce the ``best'' output of the fitness function. The first step is to randomly initialize potential solutions, and then evaluate the fitness of each potential solutions. Then while the termination criteria are unmet, a number of potential solutions are selected from the population, recombined to form children and then mutated. Selection, crossover (also called recombination), and mutation are considered the operators for the genetic algorithm. Then each child's fitness is evaluated and some number in the original population will be replaced by the child.

Our genetic algorithm uses uniform crossover with 2 parents and fitness proportionate selection. Fitness proportionate selection associates a probability of selection with a particular individual in the population \citep{hancock1994empirical}. If $fit_{t}$ is the fitness of individual $t$, the probability of selection is $$p_{t} = \frac{fit_{t}}{\sum_{j=1}^{N}fit_{j}}$$ where $N$ is the number of individuals in the population.  Fitness proportionate selection tends to be relatively slow \citep{goldberg1991comparative} and has a risk of slow convergence \citep{mitchell1997machine}. Uniform crossover allows parents to contribute individual genes to a child individual rather than sequences of genes. This eliminates positional bias \citep{mitchell1997machine} which was necessary as positional bias could interfere with certain metamorphic relations. The probability that a parent will contribute a particular gene to the child is $p_{t} = \frac{1}{a}$ where $a$ represents the number of parents. 

When designing a genetic algorithm for optimization, if there is a single objective function and no constraints, creating a fitness function is very simple. We simply use the objective function itself as a measure of fitness. It is also trivial to regularize the output. In this project we will test both stochastic and deterministic objective functions. 

In addition to implementing our own Genetic Algorithm, we contacted Strasser et.al. \citep{strasser2016fea} who allowed us to use their Factored Evolutionary Algorithms framework (FEA framework) as an additional implementation to test. 

\subsection{Differential Evolution}
Differential Evolution (DE) is another evolutionary algorithm that was designed for continuous spaces. DE centers around creating offspring by conducting crossover on parents and what is known as a ``trial vector'' \citep{storn1995differential}. A trial vector is similar to the child concept in genetic algorithms. A trial vector is generated by choosing 3 individuals -- call them $x_i(t)$, $x_2(t)$, and $x_3(t)$ -- from the population, without replacement, where $x_2$, and $x_3$ are randomly selected. The trial vector, $u_i$, is then

\begin{equation}
u_i(t) = x_i(t) + \beta (x_2(t)-x_3(t))
\end{equation}

\noindent The subtraction term generates what is known as a ``difference vector'', where the $beta$ multiplier is some user-defined positive number. Crossover then combines $x_i(t)$ and $u_i(t)$ into an offspring, which has the effect of ``pushing'' $x_i$ in a particular direction in the search space. There are multiple forms this crossover can take; we chose to implement binomial crossover, which is similar to uniform crossover in a genetic algorithm.

Because of the representation of our chromosomes, we used binomial crossover. We used a single difference vector, and to select the target vector we used random selection. As such our differential evolution algorithm could be described as $DE/random/1/binomial$.  We implemented differential evolution ourselves and used the differential evolution algorithm from the FEA framework. 

\subsection{Test Problems} 

There are a variety of established test problems used in order to assess the ability of the genetic algorithm to optimize. For the purposes of this project, we wanted problems that were continuous and scalable to more than three dimensions. Differential evolution operates almost exclusively in continuous spaces. Although the current genetic algorithm will be able to optimize in discrete and categorical spaces, we wanted problems that would also work for the differential evolution algorithm. We need problems that are scalable to more than three dimensions because most real-world problems in machine learning have feature spaces (or numbers of dimensions) much greater than three. 

We ran the genetic algorithm on a variety of problems in order to show how the metamorphic relations would change, or remain constant depending on the problem. To that end, we wanted at least one random function. We also wanted at least one function with multiple local minima, because local minima have a tendency to "trap" genetic algorithms. For this project, we selected three test problems from the literature \citep{jamil2013literature}, Ackleys function, the Quartic function, and the Rosenbrock function. Ackleys function is a continuous, deterministic, scalable function that has multiple local minima. This function fulfills our requirement for a function with multiple local minima.
\begin{multline}f(x_{1}, ..., x_{D}) = -20*e^{-0.2\sqrt{\frac{1}{D}\sum_{i=1}^{D}x_{i}^{2}}} - e^{\frac{1}{D}\sum_{i=1}^{D}cos(2\pi*x_{i})} + 20 + e
\end{multline}
\noindent The Quartic function is a continuous, scalable, stochastic function that, because of the random element, may or may not have local minima. This function fulfills our requirement for a random function. \begin{equation}f(x_{1}, ..., x_{D}) = \sum_{i=1}^{D}i*x_{i}^{4} + random[0,1)\end{equation}
\noindent The Rosenbrock function is a continuous, deterministic, scalable function with a single minimum. This function has neither a random element, nor multiple local minima. This means it is a useful function for comparisons against both the Quartic and Ackleys functions. \begin{equation}f(x_{1}, ..., x_{D}) = \sum_{i=1}^{D-1}\big(100(x_{i+1} - x_{i}^{2})^{2} + (x_{i}-1)^{2}\big)\end{equation}

\begin{table*}
\begin{center}
\begin{tabular}{| l  c  c  c  c |}
\hline
Class & LOC & \# of tests & \# of trad. tests & \# of MRs \\ \hline
Chromosome & 22 & 3 & 2 & 1 \\ \hline
Fitness Function & 14 & 1 & 0 & 1 \\ \hline
Ackleys & 17 & 7 & 5 & 2 \\ \hline
Quartic & 11 & 6 & 3 & 3 \\ \hline
Rosenbrock & 18 & 6 & 5 & 1 \\ \hline
Genetic Algorithm & 112 & 14 & 5 & 9 \\ \hline
Differential Evolution & 82 & 5 & 2 & 3 \\ \hline
Total & 271 & 42 & 22 & 20 \\ \hline
Total without DE & 189 & 37 & 20 & 17 \\ \hline
\end{tabular}
\caption{Lines of code, number of tests written, number of deterministic tests and number of tests written using metamorphic relations.}
\label{tab:GAInfo}
\end{center}
\end{table*}
The genetic algorithm we implemented is organized as follows: We used three fitness functions in this work, Ackleys, Quartic and Rosenbrock functions. The Chromosome class encapsulates a potential solution to a particular problem. The Genetic Algorithm class is the biggest class by far. It contains a list of Chromosomes which represent the population. It also contains all the operators for the Genetic Algorithm, a multitude of getters and setters, and the genetic algorithm itself, modeled on the genetic algorithm we outlined in Algorithm \ref{algo:GA}. 

The Differential Evolution class is similar to the Genetic Algorithm class. For most analyses, the Differential Evolution class was not included. 

The FEA framework is a much bigger program, containing over 5000 lines of code. For comparison, our implementation used only 271 lines of code. However, much of the FEA framework code implements other algorithms such as Particle Swarm Optimization or other test problems such as the Rastrigrin Function. The lines of code directly relating the the genetic algorithm, the differential evolution algorithm or the three fitness functions only amounts to 245 lines of code. The FEA framework can be loosely organized into the genetic algorithm, the fitness functions and the differential evolution algorithm. 

\section{Metamorphic Relations}
\label{sec:metarel}
In this section we lay out 17 metamorphic relations. Three of these relations also applied to differential evolution.

\subsection{Metamorphic Relations for Fitness Functions}
\label{subsec:unitfit}
The fitness functions are very different from the rest of the genetic algorithm because we do have the true answers for two of the functions. For Ackleys function, the minimum output is 0, and this occurs when the input is (0,0,...,0) for all dimensions. The maximum value of Ackleys function is approximately 22.3. Since there are many peaks in Ackleys function, there are many different sets of inputs that could reach this value. In two dimensions, one of those input sets is at (-21.6, 31.5) when the number of dimensions is 2. For the Rosenbrock function, the minimum output is 0. This occurs when the input is (1,1,...,1) for all dimensions. The maximum output is approximately $8.6490961*10^{7}$. These values were tested using deterministic unit tests. 

\paragraph{Relation 1.1} Unit tests for the Quartic function were tested using metamorphic relations. Since the Quartic function adds a random number $\in [0,1)$ for each dimension, we do not know what the exact value of the Quartic function will be. However, we do know that the minimum value without the random numbers would be 0. We also know that the random numbers will all be strictly less than 1. Therefore, we know that the minimum value $MV$ for the Quartic function will be $0 \leq MV < d$ where $d$ is the number of dimensions. In certain circumstances, other inputs will cause the output to be in this range. The follow up test will be $(d-1)$ 0's and the last input value is 1. This input will produce a result always larger than $(d)$ 0's. As an example, in 4 dimensions, the initial test case would be (0,0,0,0) and the follow-up test case would be (0,0,0,1). With these inputs, the output from the initial test case would always be less than the output from the follow-up test case. 

\paragraph{Relation 1.2}The maximum value the Quartic function would take if the random elements were not added is $2.68435456*\frac{d(d+1)}{2}$. This happens at (1.28, 1.28, ..., 1.28), or (-1.28, -1.28, ..., -1.28) for all dimensions. The random number generator includes 0, so the maximum value is always greater than or equal to this number. Since there are $d$ random numbers added, and the random numbers are pulled from a uniform distribution, the mean maximum value will be $2.68435456*\frac{d(d+1)}{2} + \frac{d}{2}$ and the variance is $\frac{d}{12}$. For this test, since we have fairly complete knowledge of the distribution, we could have used simple statistical tests. However, we used statistical metamorphic testing \citep{guderlei2007statistical} because it is not always possible to know the distribution of the fitness function. To perform the statistical metamorphic tests, we generated two samples, each with 20 observations, by running the Quartic function on the input (1.28, 1.28, ..., 1.28) 20 times for each sample (e.g (1.28, 1.28) for 2 dimensions) and recording the output of the function. Our null hypothesis was that the mean of the two samples would not be equal to each other. Our alternative hypothesis was that the means would be equal to each other. 

\paragraph{Relation 1.3} One of the metamorphic relations often mentioned is changing the order of the attributes \citep{xie2011testing}. For Ackleys function, this is a valid metamorphic relation that we use in the testing process. As an example, an initial input of (6.4, 2.5, 1.25) and a follow-up input of (1.25, 2.5, 1.28) will both produce an output of 13.24197384. However, for the Quartic function and the Rosenbrock function, this is invalid. Consider the following example for the Quartic function in 3 dimensions. The first input is (0.25, 0.5, 1.28). This produces a result of 9.94196993. The second input is (1.28, 0.5, 0.25). As you can see, only the order changes between the two inputs. However, the output for the second input is 4.64107331, less than half the output of the first example. 

\paragraph{Relation 1.4}Another metamorphic relation for the Quartic and Rosenbrock functions is to compare the maximum output of the fitness functions given some potential solution inside of the range of expected values, given some potential solution outside of the range of expected values, and finally the first solution again. The fitness function class will adjust the maximum output if some solution produces a higher output than is currently the case. This happens so that we do not get a fitness greater than the maximum fitness. The fitness should change between the first and the last test, and the fitness of the first test should be higher. This does not apply to the Ackleys function because the values outside the range of expected values are not necessarily higher or lower than those inside the range.  Again, we used statistical metamorphic testing in this relation. We generated 2 samples, each with 20 observations. The initial sample was generated by feeding the maximum input (e.g. (30, 30, 30) for 3 dimensions in the Rosenbrock function) into the function 20 times. Next we ran the function with a single input that was larger than the maximum input (e.g. (80, 80, 80) in 3 dimensions for the Rosenbrock function). Finally, we generated the follow-up sample by running the function on the same input as the initial sample. Our null hypothesis was that the mean of the two samples would be equal to each other, while our alternative hypothesis was that the mean of the initial sample would be greater than the mean of the follow-up sample. 

\paragraph{Relation 1.5} Since all of our fitness functions can generalize to more than three dimensions, another metamorphic relation for the fitness functions is to create a fitness function object for two dimensions and input a good solution in two dimensions, say (1,1) for the Rosenbrock function. The follow-up test would be to create a fitness function object for greater than two dimensions and input the same good solution, simply with more dimensions, e.g. (1,1,1,1) for the Rosenbrock function. The scaled fitness for the test and the follow-up test will be the same. 

\subsection{Metamorphic relations for genetic algorithm operators}
\label{subsec:unitops}
\subsubsection{Metamorphic relations for mutation}

\paragraph{Relation 2.1}Let us assume that the mutation rate is 1. Let us also assume that we have a solution with $m$ dimensions. We then run our mutation operator on our  solution. Each observation of a value in the mutated solution will be different from the corresponding un-mutated observation by some amount. This amount will be greater than or equal to 0 and less than 0.1. The average amount will be 0.05. An example input initial test case would be the randomly generated solution (3, 27, 6, 14, 30, 16, 1, 16, 10, 29) for 10 dimensions. We then run the mutation operator, with the mutation rate set to 0.1 on this solution and record the difference between the input and the output. The follow-up test case would use another randomly generated solution with the mutation operator set to 0.9 and the difference between the input and output of the mutation operator would be recorded. Our null hypothesis was that the mean difference between the input and output for the two test cases would be equal. Our alternative hypothesis was that the mean difference for the follow-up test case when the mutation rate was set to 0.9 would be greater than the mean difference for the initial test case. 

\subsubsection{Metamorphic relations for crossover}

\paragraph{Relation 2.2} The crossover operator takes as input a set of 'parent' solutions $parents$ and outputs a 'child' solution that is a combination of elements of the parent solutions. For each dimension in the solution, the probability that the parent will contribute their value for that dimension is $\frac{1}{|parents|}$. Given a set of parents that are unique (i.e. no common elements between the parents), we can determine which element of the child solution came from which parent. This relation also used statistical metamorphic testing. With parents (1,2,3,4) and (5,6,7,8), our initial test case would set the crossover rate to 0.5 and run the crossover operator 20 times, generating 20 children. The follow-up test case would set the crossover rate to 1 and again generate 20 children. We would then calculate the proportion of elements in each child that came from the first parent. Our null hypothesis was that the average proportion for the initial sample would be equal to the average proportion for the follow-up sample. Our alternative hypothesis was that the average proportion for the initial sample would be greater than the average proportion for the follow-up sample. This metamorphic relation also applies to Differential Evolution. 
\subsubsection{Metamorphic relations for selection}

Our implementation of the genetic algorithm uses fitness proportionate selection. This means that each solution's fitness is scaled by the total fitness of the population. These scaled fitnesses all add up to 1. More specifically, this algorithm uses roulette wheel selection. This means that we create a scaled fitness vector, and we select a potential solution based on these scaled fitness vectors. Smaller fitness values are more likely to be selected, but in order to maintain diversity in our population of solutions, there must be some chance that solutions with higher fitness can be selected. 

\paragraph{Relation 2.3} This metamorphic relation involves running the selection operator several times on two populations, one that contains several copies of the ideal solution, and one that does not but is identical in every other way. For an example with the Rosenbrock fitness function and 2 dimensions, our initial test case might have the individuals (2,3), (5,10), (27, 8), (17, 11), and (29, 2). The follow-up test case might have the individuals (3,4), (5,10), (17,11), (1,1), and (1,1). The average fitness for the initial test case would be worse than the average fitness for the follow-up test case. If the ideal solution is unknown, a good solution, as determined by the fitness function, could be used instead of the ideal solution. 

\subsection{System-level metamorphic relations} 
\label{subsec:wholalgo}
\paragraph{Relation 3.1} The number of generations we allow the genetic algorithm to run is a crucial parameter. If we increase the number of generations, the average fitness will improve, unless the fitness threshold is reached, and the algorithm exits early. One way to prevent the early exit is to select a test problem that has many local optima so as to prevent early convergence, such as the Ackley's function. An example initial test case is setting the number of generations to 50. An example follow-up test case would be setting the number of generations to 5000. This relation applies to Differential Evolution. 

\paragraph{Relation 3.2}Another crucial parameter is the population size. As the population size increases, the average fitness will also improve. This is because with increased individuals in the population, we increase the chance that one of those individuals will encounter the true solution. For an initial test case, we used a population size of 5. The population size for the follow-up test case was 500. The average fitness for the follow-up test case will be better (lower) than the average fitness for the initial test case. Interestingly, for Differential Evolution smaller population sizes improve average fitness more than larger population sizes (assuming some low number of iterations and/or a problem with many local minima). This could be because the decreased diversity of the population leads to a faster decrease in fitness. Alternatively, there could be an interaction between the population size and another important parameter, $\beta$ that influences the ideal size of the population. We used the same initial and follow-up test cases for differential evolution. The average fitness for the follow-up test case will be worse (larger) than the average fitness for the initial test case.

\paragraph{Relation 3.3} Finally, if we increase the threshold parameter, the average fitness will be worse, but the average number of iterations run by the algorithm will decrease. This is because when the threshold is set at a higher value the algorithm reaches the threshold relatively quickly and exits. When the threshold is set at a lower value, the algorithm continues searching until it reaches the lower value, so the fitness will be lower (better). Our initial test case was 0.5 and our follow-up test case was 0.05. The average fitness for the initial test case was greater than the average fitness for the follow-up test case. The average number of iterations run for the initial test case was less than the average number of iterations run for the follow-up test case. 

\begin{figure*}
\begin{center}
\includegraphics{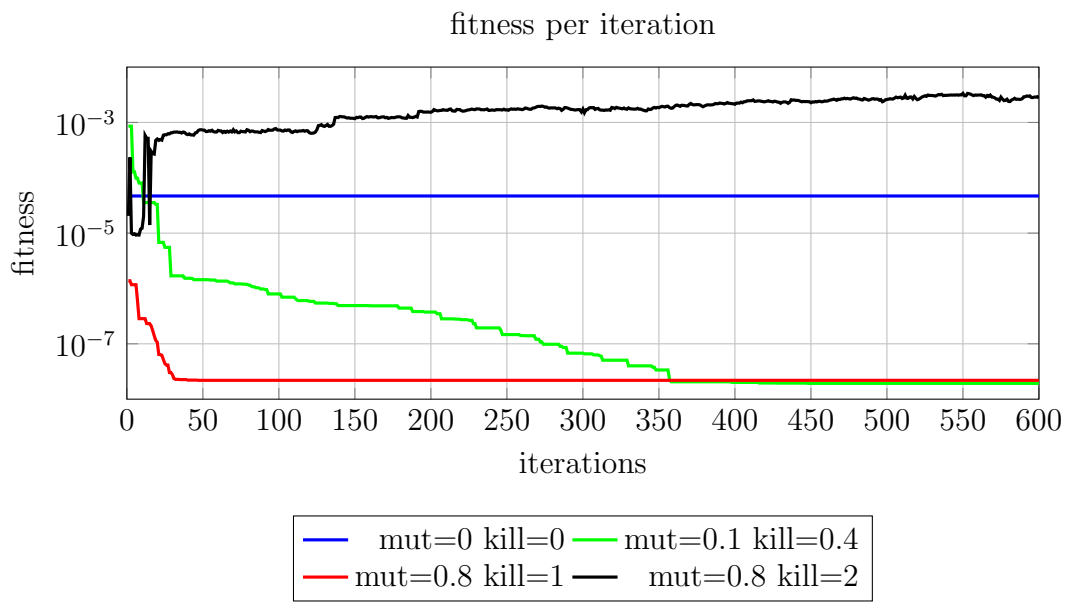}
\end{center}
\label{fig:fitness}
\caption{Fitness for the best solution at each iteration when the parameters are varied. When fitness decreases, the solution is improving.}
\end{figure*}

The other two parameters, mutation rate and replacement rate, are much more difficult to define relations for. In Differential Evolution, $\beta$ and the crossover rate can be seen as approximate substitutions for mutation and replacement rates. As you can see from Figure 1, when both of these parameters are at 0, we will see no improvement in total fitness in the population over the generations. As we increase the parameters, we will see improvements in the total fitness and the best fitness will be reached in fewer iterations. At a certain point, total fitness will oscillate. If the oscillations are small, we should still see convergence towards our ideal solution. However, these oscillations, no matter how small, will lead to a greater number of generations needed to find the ideal solution. This point where total fitness begins to oscillate is not easily identifiable and is thought to depend on the problem. We can see that in this instance, unstable oscillations occur when the mutation rate is at 0.8 and the replacement rate is at 2. When the unstable oscillations occur, overall fitness increases rather than decreasing. It is unclear how the mutation rate and replacement rate interact to speed or slow the rates of convergence towards the best solution in general. However, we can still define some metamorphic relations for mutation rate and replacement rate. 

\paragraph{Relation 3.4} When both parameters are at 0, the average fitness will be worse than if both parameters are at 0.5. This is because the final solution depends only on the randomly initialized potential solutions, rather than the changes made to those solutions. This applies to crossover and $\beta$ in Differential Evolution. 

\paragraph{Relation 3.5} On the other hand, if both parameters at 1 the average fitness does not follow the same behavior. We think this is because each time mutation happens, the solution will only change by a small amount. Recombination never adds new solutions. Thus each solution is only being changed by a small amount. It is possible that the mutation operator can be changed so that setting both parameters equal to 1 will be roughly equivalent to a random search. This relation was tested in section \ref{subsec:failure}, but was not tested in the remainder of the experiments. 

\paragraph{Relation 3.6} If we hold mutation rate constant at 0, and increase the replacement rate to 0.5, the average fitness of the best solution in the population will be better than when both mutation rate and replacement rate are at 0. Performing Relation 3.6 with mutation and replacement rate switched, however, does not produce the same results. This is because mutation happens only when a potential solution is selected, so the mutation rate depends on the selection rate. 

\paragraph{Relation 3.7} If however, we hold the recombination rate constant at some low number, say 0.1, and set the mutation rate at 0 and 0.5, for example, the average fitness for the best solution will be better when the mutation rate is at 0.5 than when the mutation rate is set at 0.

\paragraph{Relation 3.8} For this reason, the common wisdom in the genetic algorithm literature is that mutation rate should be set lower than the replacement rate \citep{deb1998understanding}. If this common wisdom holds, one metamorphic relation for mutation is to run the whole algorithm when the parameters follow this common wisdom, and then run the algorithm again when the values for the parameters are swapped. The average fitness for the best solution when the parameters follows this common wisdom should be better than when the parameters do not follow this common wisdom. However, when the values are 0.1 and 0.8, this is not the case. Much of the time, the swapped parameters (i.e. the mutation rate set higher than the replacement rate), performed better than the parameters that followed the common wisdom. This high rate of failure is one reason this test was not used in the analyses, other than in section \ref{subsec:failure}.

\paragraph{Relation 3.9} On the other hand, we can test this type of interaction between the replacement and mutation rates by using the values 0 and any other number strictly greater than 0 and less than or equal to 1. In that case, the algorithm should behave like Relation 3.6 with the parameters switched. In other words, the average best fitness for this algorithm will be no better than when both parameters are set at 0.

\subsection{Deterministic unit tests}
\label{subsec:unitdeter}
We implemented several unit tests for both the genetic algorithm and differential evolution that did not use metamorphic relations. These relations included testing initialization, selecting the best chromosome from the population, that fitness is changed when the update fitness function was called, and checking constant and known values. These were used for comparisons against the metamorphic relations. The scope of these deterministic tests is limited, but testing many parts of a genetic algorithm is difficult, if not impossible, without metamorphic testing. 

\section{Experiments}
\label{sec:exp}
In order to conduct mutation testing, we used the PIT mutation testing tool \citep{coles2012pit}. Although the PIT tool does not include source code of the mutants, it does include the line number where the mutation happened and the type of mutation (e.g. negated conditional, replaced addition with subtraction). PIT reports each mutant that survived, was killed, timed out or was not covered for each class. For each test, 182 mutants were generated. PIT uses Line coverage to assess mutants not covered by test cases. 
The PIT mutation tool generates the following types of mutants:
\begin{itemize}
\item Replaced operator ($+, -, *, /$) with another operator.
\item Changed conditional boundary.
\item Negated conditional.
\item Changed increment.
\item Mutated return value. 
\item Removed call to other function.
\end{itemize}
Unless otherwise specified by the metamorphic relation, the parameters for the genetic algorithm were as follows:
\begin{itemize}
\item $popSize = 50$.
\item $mutRate = 0.1$.
\item $killRate = 0.4$.
\item $\delta = 0$.
\item $maxGen = 1000$.
\end{itemize}

\begin{table*}
\begin{center}
\begin{tabular}{| l  c  c  c  c  c  c |}
\hline
Statements & Total & Deterministic & Total & Function & System & Total \\ 
Covered & Tests & Tests & MRs & Level MRs & Level MRs & Lines\\ \hline
Our Implementation & 243 & 87 & 169 & 136 & 165 & 194\\ \hline
FEA Framework & 125 & 119 & 125 & 124 & 125 & 245 \\\hline

\end{tabular}
\end{center}
\label{tab:lc}
\caption{Statement Coverage for each set of tests, reported as a ratio between lines of code covered over total lines of code, leaving aside the differential }
\end{table*}

\subsection{Overall results}
\label{subsec:overall}
Relation 2.3 detected a fault in our original implementation of the genetic algorithm. The fault in the selection function occurred because we were prioritizing higher fitness, rather than lower fitness. Once we had fixed this error, we generated mutation scores for all types of tests. As shown in Table \ref{tab:allResults}, the overall mutation score is 85\% for all the tests written. Several classes received a mutation score of 100\%. The vast majority of mutants that were not killed were due to there being no coverage, or tests, for the code that was changed (44 of the 59 mutants that were not killed). This is a problem that can be easily remedied with more tests. 

\begin{table*}
\begin{center}
\begin{tabular}{| c  c  c  c |}
\hline 
& All Tests & Deterministic Tests& All Relations \\ \hline
Genetic &90/117 & 20/117 & 90/117 \\ 
Algorithm & 77\% & 17\% & 77\% \\ \hline
Chromosome & 14/15 & 8/15 & 11/15 \\
& 93\% & 53\% & 73\% \\ \hline
Fitness & 9/11 & 5/11 & 7/11 \\ 
Function & 81\% & 46\% & 64\% \\ \hline
Ackleys & 13/14 & 11/14 & 12/14 \\ 
Function & 93\% & 79\% & 86\% \\ \hline
Quartic & 8/8 & 0/8 & 8/8 \\
Function & 100\% & 0\% & 100\% \\ \hline
Rosenbrock & 18/18 & 16/18 & 18/18 \\ 
Function & 100\% & 89\% & 100\% \\ \hline
Combined & 104/132 & 28/132& 101/132 \\
Gen. Algo. & 79\% & 21\% & 77\% \\ \hline
Combined & 48/51 & 32/51 & 45/51 \\  
Fit. Func. & 94\% & 63\% & 88\% \\ \hline
Total & 152/182  & 60/182 & 147/182 \\ 
 & 84\% & 33\% & 81\% \\ \hline

\end{tabular}
\caption{Mutation scores for the genetic algorithm implementation comparing deterministic tests and all tests involving metamorphic relations. We report totals by class and then totals for the all classes with Genetic Algorithm components (Genetic Algorithm and Chromosome) and all classes for fitness function components (Fitness Function, Ackleys Function, Quartic Function, and Rosenbrock Function) }
\label{tab:allResults}
\end{center}
\end{table*}

\subsection{Deterministic and Metamorphic Comparison}
\label{subsec:determr}
We next divided the tests into deterministic and metamorphic tests. All tests were run again and the mutation score was calculated for each type of test. For the deterministic tests, the mutation score was 33\%, as seen in Table \ref{tab:allResults}. This is quite low, due at least in part to not being able to test the random elements of the genetic algorithm. The mutation score for metamorphic testing was 81\%. Part of the reason for the big disparity was potentially due to the fact that we have identified two different types of metamorphic tests, function-level and system-level metamorphic relations.  

\begin{table*}
\begin{center}
\begin{tabular}{| c  c  c  c |}
\hline 
& All Relations & Function-Level & System-Level \\ \hline
Genetic & 90/117 & 48/117 & 74/117 \\ 
Algorithm & 77\% & 41\% & 64\% \\ \hline
Chromosome & 11/15 & 11/15 & 11/15\\
&73\% & 73\% & 73\% \\ \hline
Fitness & 7/11 & 7/11 & 6/11 \\ 
Function & 64\% & 64\% & 55\% \\ \hline
Ackleys & 12/14 & 12/14 & 8/14 \\ 
Function & 86\% & 86\% & 57\% \\ \hline
Quartic & 8/8 & 8/8 & 8/8 \\
Function & 100\% & 100\% & 100\%\\ \hline
Rosenbrock & 18/18 & 18/18 & 7/18\\ 
Function & 100\% & 100\% & 39\% \\ \hline
Combined & 101/132 & 59/132 & 85/132 \\
Gen. Algo. & 77\% & 45\% & 64\% \\ \hline
Combined & 45/51 & 45/51 & 29/51 \\  
Fit. Func. & 88\% & 88\% & 57\% \\ \hline
Total & 147/182 & 104/182 & 114/182\\ 
 &81\% & 57\% & 63\% \\ \hline

\end{tabular}
\caption{Mutation scores for the genetic algorithm implementation comparing function-level and system-level tests. We report totals by class and then totals for the all classes with Genetic Algorithm components (Genetic Algorithm and Chromosome) and all classes for fitness function components (Fitness Function, Ackleys Function, Quartic Function, and Rosenbrock Function) }
\label{tab:mrResults}
\end{center}
\end{table*}

\subsection{System-level and Function-level comparison}
\label{subsec:unitwholealgo}
Since the deterministic tests were implemented only at the function level and there were no system-level deterministic tests, we divided the metamorphic relations into system-level and function level tests. The mutation score for the function level metamorphic relations was 0.577, from Table \ref{tab:mrResults}. This was higher than the deterministic tests. We were originally concerned that this was due to differences in coverage of the different types of tests. However if we approximate a normalization by dividing the number of mutants killed by the lines of code covered, we see that the deterministic tests only scored 0.588, while the function level metamorphic relations obtained a score of 0.772 (the best possible score would be 1.055). This means that the deterministic tests had a lower mutation score, even when measured relative to statement coverage. The mutation score for the system-level relations was even higher than for the function-level relations, at 0.648. Most of what is driving that number is the higher coverage for the genetic algorithm class. As you can see in Table \ref{tab:mrResults}, when separated out by class, the whole algorithm relations had a mutation score of 0.638 for the Genetic Algorithm class, while the function-level relations only had a mutation score of 0.414. On every other class, the function-level relations had higher mutation scores. 

\subsection{System-level Tests with Different Fitness Functions}
\label{subsec:failure}
Finally, we examined the failure rates of the whole algorithm tests. We expected that some small number of tests would fail if the tests were run a sufficient number of times, given the nature of statistical tests. However, we suspected we were seeing too many failures for statistical likelihood. If this was due to an actual error in the genetic algorithm, we expected that the test failure would be consistent. The failures we were seeing occurred inconsistently. One potential cause of this was the fact that we were using the Ackleys function to set up the tests for several metamorphic relations. Ackleys function has many local optima. We hypothesize that if there were more failures than expected, these failures were due to the Ackleys Function getting stuck in local optima and not converging towards a global optima. 

We restricted the set of tests run to the system-level metamorphic relations, specifically to Relations 3.1, 3.2, 3.3, 3.4, 3.5, and 3.8. We ran each set of relations ten times. We then changed the fitness functions for each relation and ran the relations another ten times. For example, we would run Relation 3.1 ten times with Ackleys function, then ten times with the Quartic function, and finally ten times with the Rosenbrock function. If statistical tests were causing the failures, we expected that failure would occur less than one time for each fitness function with each relation.

\begin{table*}
\begin{center}
\begin{tabular}{| l  c  c  c  c  c  c |}
\hline 
& 3.1 & 3.2 & 3.3 & 3.4 & 3.5 & 3.8 \\ \hline
Ackleys & 7 & 0 & 0 & 0 & 2 & 8 \\ \hline
Quartic & 6 & 0 & 0 & 0 & 3 & 5 \\ \hline
Rosenbrock & 0 & 0 & 0 & 0 & 4 & 5 \\ \hline
\end{tabular}
\caption{Number of failures in ten runs for each relation when the fitness function is varied}
\label{tab:fail}
\end{center}
\end{table*}

As you can see from Table \ref{tab:fail}, only 3 relations failed at all, and all the failing relations failed much more than we would expect given a 95\% confidence interval. Relations 3.2, 3.3 and 3.4 succeeded consistently. We restricted our further tests to Relations 3.1, 3.5 and 3.8. Relation 3.1 failed only when run with Ackleys and Quartic functions. This fit our hypothesis that failing tests were getting stuck in local optima. We ran this relation 20 more times with just the Rosenbrock function and found that it did not fail in any of those runs. However, running Relations 3.5 and 3.8 with the Rosenbrock function did not alter the amount of failures seen. One option for Relation 3.5 was to increase the average alteration made to each value performed by the mutation operator, in hopes that it would improve the failure rate. However, this would require altering Relation 2.1 to match. Instead we developed Relations 3.6 and 3.7 to substitute for this relation. Relation 3.8 was the most problematic relation, also probably requiring a change to the mutation operator. We identified Relation 3.9 to replace Relation 3.8. After we replaced the faulty relations, we ran Relations 3.1, 3.2, 3.3, 3.4, 3.6, 3.7, and 3.9 another twenty times each, and saw no more failures. 

\subsection{FEA framework}
We used the same tests and relations for the FEA framework as we did for our implementation. The authors of \citep{strasser2016fea} had previously constructed 70 unit tests for the FEA framework. All but 2 of the unit tests were irrelevant to the genetic algorithm, fitness functions and differential evolution. As you can see in Table \ref{tab:gaData} the whole algorithm relations achieved the highest mutation score in the genetic algorithm classes, while the unit level relations achieved higher mutation scores for the fitness function classes. Both types of relation had higher mutation scores than the deterministic unit tests. Since more deterministic tests were implemented, but the mutation scores for the relations was higher, we can surmise that the metamorphic relations were far more effective than the deterministic unit tests. When we compared the results of our implementation to the FEA framework we found lower mutation scores for the FEA framework, as seen in Table \ref{tab:compData}. This suggests that we were possibly targeting our relations to the implementation. However, the mutation scores for the metamorphic relations were still higher than the mutation scores for the deterministic tests, suggesting that these relations are still more effective at detecting errors than deterministic tests.

\begin{table*}
\begin{center}
\begin{tabular}{| c  c  c  c  c |}
\hline 
&Initial & All Tests & Deterministic Tests & All Relations \\ \hline
Genetic &0/89 &62/89    & 35/89 & 62/89 \\ 
Algorithm & 0\% & 70\%    & 39\%  & 70\%  \\ \hline
Fitness& 24/71 &47/71     & 35/71 & 43/71 \\ 
Functions& 33\% & 66\%     & 49\%  & 60\%  \\ \hline
Total& 24/160 & 109/160     & 70/160 & 105/160  \\ 
& 15\%& 68\%     & 44\%  & 66\%  \\ \hline
\end{tabular}
\caption{Mutation scores for the FEA framework comparing initial results, deterministic tests and all tests involving metamorphic relations}
\label{tab:gaData}
\end{center}
\end{table*}

\begin{table*}
\begin{center}
\begin{tabular}{| c  c  c  c |}
\hline 
& All Relations & Function-Level & System-Level \\ \hline
Genetic & 62/89 & 37/89     & 62/89     \\ 
Algorithm & 70\%  & 42\%      & 70\%       \\ \hline
Fitness & 43/71 & 39/71     & 36/71     \\ 
Functions& 60\%  &  54\%     & 50\%       \\ \hline
Total& 105/160 & 76/160     & 98/160     \\ 
 & 66\%  &  48\%     & 61\%       \\ \hline
\end{tabular}
\caption{Mutation scores for the FEA framework comparing function-level and system-level tests.}
\label{tab:mrData}
\end{center}
\end{table*}

\begin{table*}
\begin{center}
\begin{tabular}{| c  c  c  c  c  c  c |}
\hline 
 &Initial & All & Deterministic & MRs & Function MRs & System MRs \\ \hline

FEA &	15\%   & 68\% & 44\%  & 66\%  &  48\%     & 61\%       \\ 
framework & & & & & & \\ \hline
Our & 0\% & 83\% & 33\% & 81\%  &  58\%     & 64\%       \\ 
impl. & & & & & & \\ \hline
\end{tabular}

\caption{Comparison between mutation scores for our implementation and the FEA framework}
\label{tab:compData}
\end{center}
\end{table*}

\subsection{Differential Evolution}
We ran five tests from the genetic algorithm on the differential evolution algorithm on each implementation. Two of these tests were deterministic while the remaining three were metamoprhic relations. Tests performed much better than expected on the differential evolution algorithm, given that all but one of the tests was virtually identical to the genetic algorithm tests. The mutation scores for all tests for the FEA framework implementation of differential evolutions was 95\% while for our implementation the mutation score for all tests was 86\%. This was the opposite of the genetic algorithm tests in that the FEA framework tests performed better than the tests on our implementation. Additionally, in the FEA framework, differential evolution tests outperformed our implementation of the genetic algorithm in terms of mutation score, while our implementation of the genetic algorithm slightly outperformed our implementation of differential evolution. This increase in performance is intriguing, although it probably can be attributed to differential evolution being a simpler program with fewer random elements. 

\begin{table*}
\begin{center}
\begin{tabular}{| c  c  c  c |}
\hline 
 & All & Deterministic & MRs \\ \hline

FEA & 37/39 & 19/39 & 31/39 \\ 
framework & 95\% & 49\% & 79\% \\ \hline
Our & 50/58 & 7/58 & 42/58 \\ 
impl. & 86\% & 12\% & 72\% \\ \hline
\end{tabular}

\caption{Comparison between mutation scores for our implementation and the FEA framework on the differential evolution algorithm}
\label{tab:DEcompData}
\end{center}
\end{table*}

\section{Conclusions}
\label{sec:concl}
\subsection{Contributions}
\label{subsec:contr}
In this work we identified metamorphic relations of genetic algorithms, and genetic algorithm operators. Additionally, we have identified metamorphic relations of genetic algorithms that translate to differential evolution, and may be more illustrative in differential evolution than in genetic algorithms. We defined 17 metamorphic relations, five for the fitness functions, three for the genetic algorithm operators and nine for the whole algorithm. 

We compared the metamorphic relations to the deterministic unit tests for the genetic algorithm. We found that metamorphic relations for our implementation had a mutation score of 81\% while traditional deterministic unit tests had a mutation score of 33\%.  We also compared the function-level metamorphic relations to the system-level metamorphic relations on the genetic algorithm for our implementation. Function-level metamorphic relations had a mutation score of 57\% while system-level metamorphic relations had a mutation score of 63\%. For the FEA framework, the mutation score for the metamorphic relations was 66\% while the mutation score for the deterministic tests was 44\%. When comparing system-level and function-level metamorphic relations for the FEA framework, the mutation score for system-level metamorphic relations was 61\%. The mutation score for the function-level metamorphic relations was 48\%. Additionally, we modified two relations that failed more often than was statistically likely on both implementations of the genetic algorithm when no fault was present. Finally, we used 5 tests, two deterministic unit tests and three metamorphic relations, on the differential evolution algorithm. The mutation score for the metamorphic relations was 79\% for the FEA framework and 72\% for our implementation. The mutation score for the deterministic tests was 49\% for the FEA framework and 12\% for our implementation.

These comparisons demonstrated the effectiveness of the metamorphic testing approach when testing genetic algorithms. We assessed the mutation score relative to the statement coverage, and found that function-level metamorphic relations performed better than function-level deterministic tests, despite there being more deterministic tests. This result was consistent across both genetic algorithm implementations we tested. We examined the failure rates of the system-level relations when initialized with various fitness functions. We found one relation that only performed well when paired with a particular fitness function. Additionally, we found two relations that do not perform well no matter the fitness function. We developed two new relations to replace these problematic ones. 

\subsection{Future Work}
\label{subsec:future}
Future work for metamorphic testing on genetic algorithms would include identifying individual metamorphic relations that are either more generalizable, or kill more mutants than other metamorphic relations. We also plan to test these relations on different types of operators and different purposes for a genetic algorithm. We plan to develop relations for other operators, purposes, or algorithms. We can then also test those relations on this genetic algorithm implementation.  We would like to identify types of mutants that survive more often, in order to identify metamorphic relations that are able to target these mutants, or show that these mutants are more likely to be equivalent. Finally, any future work would have to include the identification of more metamorphic relations, and the implementation of more tests. 


\bibliographystyle{model1-num-names}
\bibliography{references.bib}

\end{document}